\documentclass[twocolumn,prl,preprintnumbers,superscriptaddress,showpacs]{revtex4}

\usepackage{graphicx}
\usepackage{amsmath,amsfonts,amssymb,amsxtra}
\usepackage{textcomp}
\usepackage{amsmath}
\usepackage{exscale}

\makeatletter
\DeclareRobustCommand{\chemical}[1]{%
  {\(\m@th
   \edef\resetfontdimens{\noexpand\)%
       \fontdimen16\textfont2=\the\fontdimen16\textfont2
       \fontdimen17\textfont2=\the\fontdimen17\textfont2\relax}%
   \fontdimen16\textfont2=2.7pt \fontdimen17\textfont2=2.7pt
   \mathrm{#1}%
   \resetfontdimens}}
\DeclareRobustCommand{\bchemical}[1]{%
  {\(\m@th
   \edef\resetfontdimens{\noexpand\)%
       \fontdimen16\textfont2=\the\fontdimen16\textfont2
       \fontdimen17\textfont2=\the\fontdimen17\textfont2\relax}%
   \fontdimen16\textfont2=2.7pt \fontdimen17\textfont2=2.7pt
   \mathbf{#1}%
   \resetfontdimens}}
\makeatother

\newcommand{\lmo}{\chemical{LaMnO_3}}

\newcommand{\tmo}{\chemical{TbMnO_3}}
\newcommand{\jfm}{\chemical{J_{FM}}}
\newcommand{\jafm}{\chemical{J_{AFM}}}
\newcommand{\jnnn}{\chemical{J_{nnn}}}
\newcommand{\eg}{\chemical{e_g}}
\newcommand{\tc}{\chemical{T_C}}
\newcommand{\tn}{\chemical{T_N}}

\newcommand{\vq}{\chemical{{\bf q}}}

\newcommand{\vQ}{\chemical{{\bf Q}}}

\newcommand{\kommentar}[1]{}

\begin{document}

\textheight 24.4 true cm

\title{Magnetic excitations in multiferroic TbMnO$_3$ }

\author{D. Senff}
\affiliation{II. Physikalisches Institut, Universit\"at zu K\"oln, Z\"ulpicher Str. 77, D-50937 K\"oln, Germany}

\author{P. Link}
\thanks{Spektrometer PANDA, Institut f\"ur Festk\"orperphysik, TU
Dresden}
\affiliation{ Forschungsneutronenquelle Heinz
Maier-Leibnitz (FRM II), TU M\"unchen, Lichtenbergstr. 1, 85747
Garching , Germany}

\author{K. Hradil}
\affiliation{ Institut f\"ur Physikalische Chemie, Universit\"at
G\"ottingen, Tammanstr. 6, 37077 G\"ottingen , Germany}

\author{A. Hiess}
\affiliation{Institut Laue-Langevin, Bo\^ite Postale 156, 38042
Grenoble Cedex 9, France}

\author{L.P. Regnault}
\affiliation{CEA-Grenoble, DRFMC-SPSMS-MDN, 17 rue des Martyrs,
38054 Grenoble Cedex 9, France}

\author{Y. Sidis}
\affiliation{ Laboratoire L\'eon Brillouin, C.E.A./C.N.R.S.,
F-91191 Gif-sur-Yvette Cedex, France}

\author{N. Aliouane}
\affiliation{Hahn-Meitner-Institut, Glienicker Str. 100,  14109
Berlin, Germany}

\author{D.N.  Argyriou}
\affiliation{Hahn-Meitner-Institut, Glienicker Str. 100,  14109
Berlin, Germany}

\author{M. Braden}
\email{braden@ph2.uni-koeln.de}%
\affiliation{II. Physikalisches Institut, Universit\"at zu K\"oln, Z\"ulpicher Str. 77, D-50937 K\"oln, Germany}

\date{\today, \textbf{preprint}}


\begin{abstract}

The magnetic excitations in multiferroic TbMnO$_3$ have been
studied by inelastic neutron scattering in the spiral and
sinusoidally ordered phases. At the incommensurate magnetic zone
center of the spiral  phase, we find three low-lying magnons
whose character has been fully determined using
neutron-polarization analysis. The excitation at the lowest
energy is the sliding mode of the spiral, and two modes at 1.1 and
2.5\ meV correspond to rotations of the spiral rotation plane.
These latter modes are expected to couple to the electric
polarization. The 2.5\ meV-mode is in perfect agreement with
recent infra-red-spectroscopy data giving strong support to its
interpretation as an hybridized phonon-magnon excitation.

\end{abstract}

\maketitle

The perovskite \tmo \ \cite{kimura-nat}  is a key material for the
new class of multiferroic transition metal oxides
\cite{goto,hur,lawes}, since the polarization in the
ferroelectric phase is sizeable and since the magnetoelectric
coupling is remarkably large \cite{kimura-nat,kimura-prb}. The
antiferro-type ordering of the single-occupied \eg -orbitals
implies a ferromagnetic nearest neighbor (nn) interaction \jfm \
in \lmo \ as well as in \tmo \ \cite{kimura-2003}, but  in \tmo ,
a large $c$-axis octahedron rotation \cite{struc-rmo} yields a
sizeable overlap of the \eg -orbitals of next-nearest neighbor
sites (nnn) along $b$ rendering the associated magnetic
interaction strongly antiferromagnetic, \jnnn \
\cite{kimura-2003}. The well-defined ferromagnetic order in the
$a,b$ planes of \lmo , therefore, becomes strongly frustrated in
\tmo \ giving rise to the complex magnetic ordering which finally
causes the multiferroic behavior.

\begin{figure}
\includegraphics*[width=0.48\textwidth]{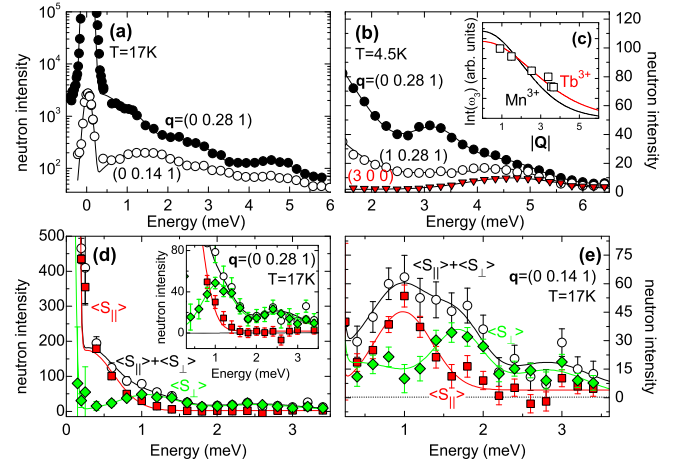}
\caption{(color online) Raw-data scans  at the incommensurate
zone center and at the $b$-axis zone boundary in the spiral phase
(PANDA and 4F, $k_f$=1.5\AA $^{-1}$) a); \vQ -dependence of the
signal  b), the insert c) compares the suppression of the 4.5\
meV feature with the expectation due to a Tb magnetic form factor
(PUMA, $k_i$=2.66\AA $^{-1}$).  Full polarization analysis of the
magnetic excitations at the incommensurate zone center and at the
zone boundary (IN14, $k_f$=1.5\AA $^{-1}$) d) and e). } \label{f1}
\end{figure}

At \tn =42\ K, the Mn spins in \tmo \ order in a longitudinal
spin-density wave (SDW) with a wave-vector of \vq =(0,0.28,0) in
reduced lattice units of the $Pbnm$-structure
\cite{quezel,kenzelmann}. Upon further cooling, the modulation
vector changes slightly until at \tc =28\ K a second transition
into a spiral phase occurs \cite{kenzelmann,kimura-nat}. Here,
the magnetic order corresponds to an elliptic cycloid still
modulated along the $b$-direction but with the spins rotating
around the $a$-axis \cite{kenzelmann}. Associated with this SDW to
spiral transition, the spontaneous electric polarization parallel
to the $c$-axis appears.

The appearance of the electric polarization at the transition
into the spiral phase was microscopically explained by Katsura et
al. \cite{katsura}. This idea was followed by studies of  the
phenomenology \cite{mostovoy} and its extension to the lattice
interaction \cite{sergienko}. Due to the non-collinear spin
arrangement in the spiral phase, the inverse of the
Dzyaloshinski-Moriya interaction implies an uniform displacement
with an electric polarization given by:

$$ {\bf{P}} \propto  {\bf{r_{ij}}} \times ({\bf{S_i}} \times {\bf{S_j}}) ~~~~~~(1),$$

for a pair of spins ${\bf S_i}$,${\bf S_j}$ with a distance vector
${\bf r_{ij}}$.  Considering the mechanism in equation (1) there
is a close coupling between the dielectric properties and the
magnetic excitations which should lead to hybridized phonon-magnon
excitations. First evidence for such mixed excitations was
recently obtained in Infra-Red (IR) optical-spectroscopy
measurements \cite{pimenov},however, their role in the
multiferroic behavior is unclear as their magnetic counterparts
have been little explored thus far \cite{kajimoto-j}.

\begin{figure}
  \includegraphics[width=0.44\textwidth]{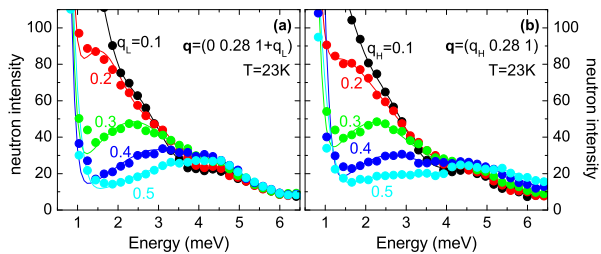}
  \caption{Raw-data scans to determine the magnon dispersion in \tmo \ along $a$ and $c$
  directions starting at the incommensurate zone center (0,0.28,1), (PUMA, $k_i$=2.662\AA $^{-1}$). }
  \label{fig2}
\end{figure}

In this letter we present the results of inelastic neutron
scattering (INS) experiments  to unravel the magnetic excitations
in \tmo , in particular the low-energy modes relevant for the
multiferroic behavior whose characters were determined  using
longitudinal polarization analysis. The excellent agreement with
the IR studies strengthens the evidence for mixed excitations.

A $\sim$900 mm$^3$ large untwinned single crystal of \tmo \ was
grown using an image furnace \cite{aliouane}. The crystal exhibits
the magnetic transitions at 42 and 28\ K in agreement with
previous studies \cite{kimura-nat,kenzelmann}. INS experiments
were performed on various triple-axis spectrometers : on 4F (cold
neutrons, Laboratoire L\'eon Brillouin), on PANDA (cold neutrons)
and PUMA (thermal beam, both at the Forschungsreaktor M\"unchen
II) and using full polarization analysis with the CRYOPAD setup
on IN14 (cold neutrons, Institut Laue Langevin). Measurements
were taken in the ferroelectric (T=17 and 23\ K), in the
paraelectric SDW and in the paramagnetic phases.

Fig. 1 shows scans taken at the incommensurate zone center \vQ
=(0,0.28,1) and at the  zone boundary in $b$ direction in the
spiral phase at T=17\ K.  Typical scans to determine the
dispersion along the $a$ and $c$-directions are shown in Fig. 2;
along these directions the frequencies rapidly increase. In all
scans we find an excitation at 4.5\ meV which in agreement with
the interpretation in reference \cite{kajimoto-j} can be
identified as a Tb crystal-field excitation, since it does not
disperse and since it exhibits a weaker dependence on the length
of the scattering vector, $\vert {\bf Q}\vert$, compared to the
other contributions. The $\vert {\bf Q}\vert$-dependence,
furthermore, is in good agreement with the Tb magnetic form
factor, see Fig. 1 c).

Fig. 3 shows the magnon dispersion of \tmo \ obtained in the
spiral (T=23\  and 17\ K) and in the paraelectric SDW phase
(T=32\ K). The dotted lines in Fig. 3 denote the magnon
dispersion measured in \lmo \ \cite{moussa}. Only along the
antiferromagnetic $c$-direction, \lmo \ and \tmo \ exhibit a
qualitatively similar dispersion. Along the $a,b$ plane, the
dispersion is essentially flattened in \tmo . The upper cut-off
energy in \tmo \ amounts to $\sim$8\ meV, whereas magnon energies
in \lmo \ extend up to 33\ meV \cite{moussa}. The pronounced
softening of the magnon frequencies for wave vectors along the
ferromagnetic planes reflects the frustrating nnn interactions and
the weakened nearest neighbor interaction \cite{kimura-2003}. The
magnons in \tmo \ are broadened, and we find evidence for a
splitting of excitations even in the SDW phase, which, however,
requires further studies. Tentatively, we may describe the
dispersion along $a$ and $c$ in a S=2 Heisenberg model with a
single magnon branch including single-ion anisotropy, D. Thereby,
we obtain the interaction parameters \jfm =0.15(1)meV (nearest
neighbors along the $a,b$ planes) and \jafm =-0.31(2)meV
(neighbors along $c$) and D=0.09(1)meV. These values should be
compared to the values of \jfm =0.83\ meV and \jafm =-0.58\ meV
observed in \lmo \ \cite{moussa} . In the spiral phase, mode
splitting is observed throughout the Brillouin zone and a more
sophisticated model to fully describe the spin-excitation
dispersion is needed.

Concerning the high-energy part of the dispersion, our data agree
qualitatively with the study of the magnon dispersion by Kajimoto
et al. \cite{kajimoto-j}, but the splitting in the excitations
has not been analyzed in reference \cite{kajimoto-j} probably due
to the fact that this group did not analyze the dispersion
starting at the incommensurate zone center. In addition, we have
explored in detail the low-energy part of the dispersion at the
incommensurate zone center and the soft branches along the
$b$-direction. As seen in Fig. 1 and in the dispersion along the
$b$ direction shown in Fig. 3 a), there are actually three
low-energy branches emerging out of \vQ =(0,0.28,1) . In view of
the magnetoelectric effect, these are the most relevant modes.
The unpolarized data in Fig. 1 a)  show the three contributions
(plus the Tb crystal field at 4.5\ meV), but this type of data
does not allow us to determine unambiguously the polarization of
the different contributions in spite of the many different
Brillouin zones explored.

We have employed longitudinal polarization analysis yielding
important additional information : in the spin-flip scattering
channel only those excitations are detected whose polarization is
perpendicular to the chosen neutron-polarization axis
\cite{polar}. We have performed the experiment with the $a$
direction vertical to the scattering plane using the
CRYOPAD-device on IN14. The setup yielded a remarkable precision
with a flipping ratio of {$I^{NSF}:I^{SF}=35$} and an accuracy of
about {$2\%$} in the transverse polarization terms. With the three
neutron-polarization directions : parallel to \vQ , (x), parallel
to $a$, (z), and perpendicular to both  \vQ \ and $a$, (y), one
may distinguish between excitations polarized perpendicular to
the spiral plane in \tmo \ (i.e. perpendicular to the $b,c$
spiral plane, $S_{\perp}$) and those polarized within the spiral
plane (i.e. in the $b,c$ plane, $S_{\parallel}$). In the
spin-flip (x)-channel the sum of both magnetic contributions
appear, whereas the spin-flip (y) and (z) channels measure only
the $S_{\perp}$ and the $S_{\parallel}$ contribution,
respectively \cite{polar}. By subtracting the intensities obtained
in different channels one directly obtains $S_{\perp}$ and
$S_{\parallel}$ without any background assumption. The two
contributions $S_{\perp}$ and $S_{\parallel}$ are separated in
the lower part of Fig. 1. At the incommensurate zone center, the
excitation with the lowest energy of 0.2(1)\ meV is $b,c$
polarized. Since one may exclude a longitudinal spin excitation
in this high-moment ordered structure, this mode must be
attributed to the transverse magnon polarized in the spiral
plane. This mode is the sliding mode of the modulated magnetic
structure its polarization scheme is shown in Fig 3d)(iii). It has
no impact on equation (1) describing the magnetoelectric coupling.
Therefore, it should not be relevant for the multiferroic
behavior. Since the spiral in \tmo \ is not perfectly circular an
anisotropy term can explain a finite energy of the sliding mode
in addition to pinning effects.

\begin{figure}
  \includegraphics[width=0.47\textwidth]{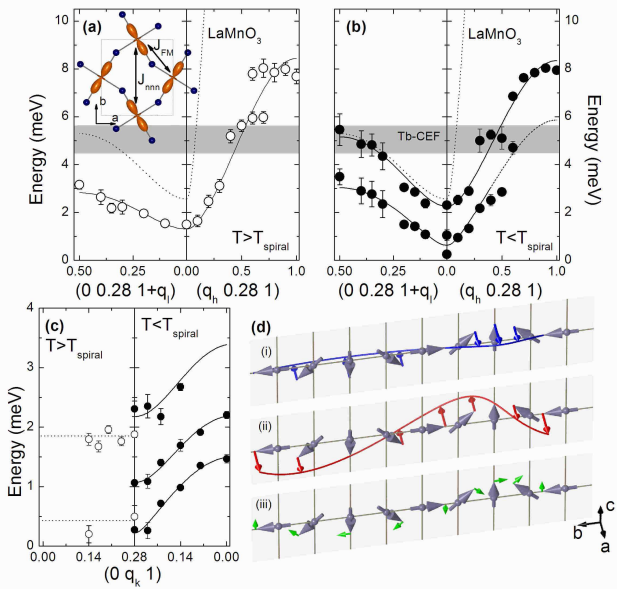}
  \caption{ Dispersion of the spin-wave excitations in  \tmo  \ along the
  $a$ and the $c$ directions of the $Pbnm$ lattice in the paraelectric
  sinusoidal phase a) and in the ferroelectric spiral phase b) .
  Dispersion along the modulation vector in the spiral and in the SDW phase c).
Polarization of the three low-energy modes at the incommensurate
zone center d) :  the two $a$-polarized modes, (i) and (ii), and
the sliding mode of the spiral, (iii), see text. }
  \label{fig3}
\end{figure}

In the $S_{\perp}$ channel, there are two contributions at
energies of 1.08(8) and 2.46(9)\ meV. The low-energy excitation
of a multiferroic system where ferroelectricity originates from
the inverse Dzyaloshinski Moriya interaction has been discussed
recently by Katsura et al. \cite{katsura-new}. Besides the
sliding mode and the phonons, there are indeed two magnetic
excitations with a spin polarization perpendicular to the spiral
plane ($b,c$ plane in the case of \tmo ) corresponding to
$S_{\perp}$. These modes are illustrated in  Fig. 3d) i) and ii).
For simplicity, we assume that the phase shift between the $b$
and the $c$ components of the spiral in \tmo \ is exactly $\pi
/2$ and that the two coefficients are of equal size, i.e. the
spiral is circular. Then, one may describe the magnetic order by :

$$ {\bf S _i}=S\cdot \cos({\bf q_{spiral}}\cdot{\bf R_i})\cdot{\bf e_y} +
S\cdot \sin({\bf q_{spiral}}\cdot{\bf R_i})\cdot{\bf
e_z}~~~~~~~~~(2),$$

where ${\bf R_i}$ is the position vector of the spin ${\bf S _i}$.
An $a$-polarized magnetic excitation at the incommensurate zone
center necessarily possesses the same modulation along $b$,
however the oscillating part can be in-phase either with the $b$
component (cosine term in equation (2), see Fig. 3d) (ii)\ ) or
with the $c$ component (sine term in equation (2), see Fig. 3d)
(i)\ ). The arising polarization patterns are distinct : in the
first case the spiral plane rotates around the $c$ direction and
it rotates around $b$ in the second case. Correspondingly, the
cross product ${\bf{S_i}} \times {\bf{S_j}}$ between neighboring
spins rotates around the $c$ direction and around the $b$
direction in these cases. Since the modulation vector or ${\bf
r_{ij}}$ in equation (1) remain always parallel $b$, the rotation
of the ${\bf{S_i}} \times {\bf{S_j}}$ cross product through the
magnon has a strong linear coupling to the polarization in the
second case, which we label $\omega _-$ following reference
\cite{katsura-new}. The $\omega _-$-excitation is the Goldstone
boson of the multiferroic phase, it may shift to finite energy
due to anisotropy terms. It is a magnon-phonon hybridization with
the phonon describing the electronic polarization along the $a$
direction. The other transverse magnon couples only
quadratically, and thus weakly, to the polarization. It weakly
modulates the static polarization parallel to $c$. Both magnons
posses finite frequencies due to anisotropy effects.

Recently, Pimenov et al. \cite{pimenov} have reported the
observation of a low-energy excitation in IR spectroscopy studies
of GdMnO$_3$ and \tmo \ which could be fully suppressed by
applying an external magnetic field. They interpreted this
excitation as an electromagnon, the hybridized magnetic and
phononic excitation. This result together with the analysis given
in reference \cite{katsura-new} agree perfectly with our INS
study of the magnetic excitations. The electromagnon feature seen
by Pimenov et al. at 20cm$^{-1}$=2.48\ meV agrees with one of the
low-lying magnetic excitations in the $S_{\perp}$ contribution.
We, therefore, tentatively attribute the 2.46 \ meV
neutron-scattering excitation to the $\omega _-$ Goldstone boson.
According to the analysis given above, this mode should couple to
the polarization in the $a$-direction, which is indeed the
direction of electric field in the IR experiment sensing the
lattice part of the excitation. In \tmo \ the spiral order might
be more complex than the ideal circular cycloid considered in
equation (2); therefore the characters of the two
$S_{\perp}$-polarized magnetic excitations might weakly mix
rendering both IR-active. It is interesting to note that also the
energy of the lower $S_{\perp}$ mode agrees perfectly with the -
though weaker - feature in the IR response at 1.24\ meV in
reference \cite{pimenov}. We emphasize, that associating the 2.46
\ meV neutron-scattering excitation with the $\omega _-$-mode
only bases on the comparison with the IR results; actually one
might expect this mode to exhibit a lower energy since it
describes the flip of the spiral into the $a,b$-plane which seems
to occur in \tmo \ under field as well as in related ReMnO$_3$
\cite{goto,aliouane,kimura-2003}.

\begin{figure}
\includegraphics[width=0.43\textwidth]{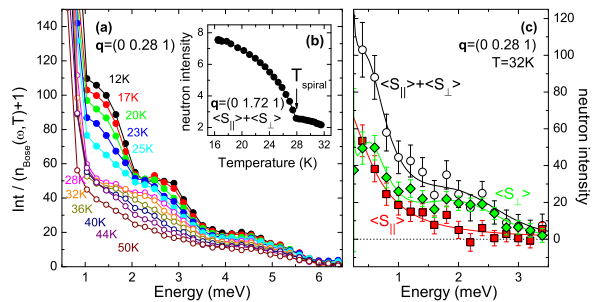}
\caption{(color online) Temperature dependence of the magnetic
scattering at the incommensurate zone center determined with
unpolarized neutrons (PUMA, $k_i$=2.662 \AA $^{-1}$ constant) a)
and with polarized neutrons (IN14, $k_f$=1.5 \AA $^{-1}$ constant)
c); the inset b) shows the magnetic elastic scattering at \vQ
=(0,1.72,1) as function of temperature. } \label{f4}
\end{figure}

Along the $b$ direction the three low-energy branches can be
followed till the incommensurate zone boundary and even beyond in
the spiral phase. The dispersion unambiguously shows that
anisotropy terms play a dominant role in \tmo ; they appear to
determine the sequence of magnetic phases as function of
temperature and magnetic field. Introducing the frustrating \jnnn
\ interaction into the non-frustrated magnon dispersion along
$b$, see insert in Fig. 3a), one may understand the $b$-dispersion
in the incommensurate phase.  The frustration generates a
dispersion minimum whose position as well as the analysis of the
exchange energy in the ordered state allow one to estimate the
interaction ratio to be of the order of \jnnn / \jfm =0.8. The
finite energies of both the $\omega _-$ feature and the other
$a$-polarized zone-center magnetic excitation arise from either a
single-ion anisotropy, from exchange anisotropy or from the
electron-lattice coupling.
More experimental and theoretical efforts are, however, needed in
order to fully characterize the aspects arising from folding the
magnon dispersion into the incommensurate ordering.

The inset of Fig. 4 shows the spin-flip signal at the (0,1.72,1)
magnetic Bragg reflection which essentially depends on the
$c$-component of the ordered moments; its temperature dependence
clearly shows the drastic increase of the $c$-component in the
spiral phase \cite{kenzelmann}. The temperature dependence of the
spectra obtained at the incommensurate zone center, see Fig. 4,
show that the two well-separated $a$-polarized excitations in the
spiral phase merge into a broad signal upon heating into the SDW
phase. In analogy to the discussion of the low-lying excitations
in the spiral phase, one may expect several contributions in the
SDW phase as well. Single-ion anisotropy causes a finite energy
for the transverse magnetic excitations and will yield a splitting
for polarization parallel $a$ and $c$. Furthermore, there will be
a separation due to the phase of the oscillating part which may be
either in-phase with the SDW or shifted by $\pi /2$. The in-phase
modes are irrelevant for the magnetoelectric coupling at least
within the Dzyaloshinski-Moriya scenario of equation (1) since
they lead to a  collinear spin structure. In contrast, the two
$\pi /2$-phase modes may couple with electric polarization even
in the paraelectric SDW phase for both magnetic polarizations.
Besides the broadening of the neutron signal in the SDW phase,
there is a temperature driven change in the spectral weights for
the different polarizations. There is still a low-lying
$S_{\parallel}$ contribution in the SDW phase in agreement with
the character of the transition into the spiral phase where
ordered moments are displaced parallel to the $c$ direction. This
low-lying $S_{\parallel}$ signal can be considered as the
quasi-soft mode associated with the SDW to spiral transition. This
mode should be strongly coupled to the polarization along $c$. The
total spectral weight of the $a$- or $S_{\perp}$-polarized
contributions, however, seems not to be focused in a single mode
but extends to higher energies. Comparing the response in the
$S_{\perp}$-channel in both phases, one may conclude that the
$\omega _-$ mode is little changing across the transition,
whereas the other $S_{\perp}$-contribution softens in the SDW
phase. These findings agree once more with the IR studies, where
the electromagnon for fields along $a$ persists into the
paraelectric SDW phase.

In conclusion the INS studies of the magnetic excitations in
multiferroic \tmo with and without polarization analysis allow us
to identify the frequencies and character of the low-energy modes
associated with the magnetoelectric coupling. The good agreement
of the observed frequencies with those of recent IR studies give
strong support to the interpretation as a magnon-phonon hybridized
electromagnon mode.

Work at Universit\"at zu K\"oln was supported by the Deutsche
Forschungsgemeinschaft through the Sonderforschungsbereich 608. We
thank N. Nagaosa and M. Mostovoy  for interesting discussions.

\begin{thebibliography}{28}



\bibitem{kimura-nat} T. Kimura et al., Nature (London) {\bf 426}, 55 (2003).

\bibitem{goto} T. Goto et al., Phys. Rev. Lett. {\bf 92}, 257201 (2004).

\bibitem{hur} N. Hur et al., Nature (London) {\bf 430}, 541 (2004).

\bibitem{lawes} G. Lawes et al.,  Phys. Rev. Lett. {\bf   93}, 247201  (2004).

\bibitem{kimura-prb} T. Kimura et al., Phys. Rev. B {\bf 71}, 184441 (2005).


\bibitem{kimura-2003} Kimura et al., Phys. Rev. B {\bf 68}, 060403(R) (2003).

\bibitem{struc-rmo} J. Blasco et al., Phys. Rev. B {\bf 62}, 5609 (2000).

\bibitem{quezel} S. Quezel et al., Physica B \& C {\bf 86}, 916 (1977).

\bibitem{kenzelmann} M. Kenzelmann et al., Phys. Rev. Lett. {\bf 95}, 087206 (2005).

\bibitem{katsura} H. Katsura et al., Phys. Rev. Lett. {\bf 95}, 057205 (2005).

\bibitem{mostovoy} M. Mostovoy, Phys. Rev. Lett. {\bf 96}, 067601 (2006).

\bibitem{sergienko} I.A. Sergienko and E. Dagotto, Phys. Rev. B {\bf 73}, 094434 (2006).

\bibitem{pimenov} A. Pimenov et al., Nature physics {\bf 2}, 97 (2006).

\bibitem{kajimoto-j} R. Kajimoto et al., J. Phys. Soc. Jpn. {\bf 74}, 2430 (2005).

\bibitem{aliouane} N. Aliouane et al., Phys. Rev. B {\bf  73},  020102 (2006).

\bibitem{moussa} F. Moussa et al., Phys. Rev. B {\bf 54}, 15149 (1996).

\bibitem{polar}R.M. Moon et al., Phys. Rev. {\bf 181}, 920 (1969).

\bibitem{katsura-new} H. Katsura et al., condmat/0602547 (2006).


\end{thebibliography}

\end{document}